# Brain Age Estimation with a Greedy Dual-Stream Model for Limited Datasets

Iman Kianian[1], Hedieh Sajedi[1]

[1]Department of Computer Science, School of Mathematics, Statistics and Computer Science, College of Science, University of Tehran, Tehran, Iran

Iman Kianian: kianian@ut.ac.ir

Hedieh Sajedi: hhsajedi@ut.ac.ir

**Abstract**

Brain age estimation involves predicting an individual's biological age from their brain images, which offers valuable insights into the aging process and the progression of neurodegenerative diseases. Conducting large-scale datasets for medical image analysis is a challenging and time-consuming task. Existing approaches mostly depend on large datasets, which are hard to come by and expensive. These approaches also require sophisticated, resource-intensive models with a large number of parameters, necessitating a considerable amount of processing power. As a result, there is a vital need to develop innovative methods that can achieve robust performance with limited datasets and efficient use of computational resources. This paper proposes a novel slice-based dual-stream method called GDSM (Greedy Dual-Stream Model) for brain age estimation. This method addresses the limitations of large dataset requirements and computational resource intensiveness. The proposed method incorporates local and global aspects of the brain, thereby refining the focus on specific target regions. The approach employs four backbones to predict ages based on local and global features, complemented by a final model for age correction. Our method demonstrates a Mean Absolute Error (MAE) of 3.25 years on the IBID's test set, which only contains 289 subjects. To demonstrate the robustness of our approach for any small dataset, we analyzed the proposed method with the IXI dataset and achieved an MAE of 4.18 years on the IXI's test set. By leveraging dual-stream and greedy strategies, this approach achieves efficiency and robust performance, making it comparable with other state-of-the-art methods. The code for the GDSM model is available at https://github.com/iman2693/GDSM.

## 1 Introduction

Brain ageing refers to the gradual changes in the brain's structure and functionality as an individual ages [1, 2]. These changes can include decreased brain volume, changes in neurotransmitter levels, and changes in cerebral blood flow [3]. Cognitive abilities may also deteriorate as people age. However, everyone ages differently, and these changes can be influenced by genetics, lifestyle, and general health. Maintaining a healthy lifestyle, exercising regularly, and controlling risk factors can help promote healthy brain ageing and potentially slow cognitive loss [4]. Regular medical examinations and cognitive evaluations can help detect serious brain health issues [5].

Brain age estimation is a computational method using neuroimaging and machine learning algorithms [6, 7] to determine an individual's biological age based on structural and functional brain traits [8]. This method offers insights into brain ageing, early detection of neurodegenerative diseases, and personalized cognitive well-being interventions [9]. It also contributes to understanding individual variances in brain health and age-related changes [10]. The term 'brain age gap' denotes the discrepancy between a person's estimated brain age and their chronological age [1, 11]. A positive gap indicates a faster aging process, potentially indicating cognitive decline or a higher risk of age-related illnesses like dementia and Alzheimer's disease [1, 12]. Conversely, a negative gap suggests a slower aging process and increased cognitive resilience [12]. Detecting a positive gap can help identify and prevent age-related disorders, enabling healthcare professionals to implement preventive measures and personalized interventions [11, 13]. Conversely, a negative gap may indicate resilience in the aging process, allowing for strategies to improve cognitive well-being and potentially reduce neurological disorders [14]. Understanding and leveraging the brain age gap can lead to more effective disease prediction, intervention, and personalized healthcare approaches.

Brain age estimation is a complex process that involves various methods, including EEG (Electroencephalography) signals [15] and MRI (Magnetic Resonance Imaging) images [16, 17]. MRI offers high-resolution structural and functional information, making it a preferred choice for accurate age prediction. The use of MRI dates back to the early 2000s when researchers applied machine learning techniques to neuroimaging data [18]. The basic methods involve extracting structural and functional features from the brain, such as gray matter volume and white matter integrity [19], and feeding these into machine learning algorithms trained on known chronological ages. However, these methods may not capture the intricate details of brain images effectively. The introduction of Convolutional Neural Network (CNN) models has significantly improved the power of feature extraction, allowing CNNs to capture spatial patterns and relationships within neuroimaging data, enhancing the model's ability to extract relevant features from MRI images [8]. This has led to improved accuracy and robustness in predicting brain age, providing more nuanced insights into age-related changes in the brain.

Further pushing the boundaries of neuroimaging research, Transformer models, originally designed for natural language processing [20], are being used in neuroimaging to improve brain age estimation capabilities [21, 22]. These models can learn complex relationships between brain regions and their dynamic interactions over time, providing a more comprehensive understanding of the aging process. The self-attention mechanism in transformers allows for effective modeling of local and global spatial dependencies in brain images, potentially leading to more accurate diagnostics and tailored interventions for age-related neurological conditions.

However, Complex models like deep 3D CNNs or transformers on small datasets can lead to overfitting issues, while simpler machine learning approaches may result in underfitting. To avoid these issues, a balance must be struck between harnessing the power of CNNs or transformers while mitigating overfitting risks, especially in limited data availability [23]. This ensures the model captures important features without succumbing to the limitations of a small dataset, enhancing performance on diverse and unseen data points. In medical imaging tasks like brain age estimation, models must navigate data scarcity while maintaining high performance.

In this study, we designed two experiments to show that the proposed approach is robust to the small datasets. Furthermore, these experiments show that the proposed approach can be used for the large-scale datasets too. The main characteristics of our proposed method are as follows:

**A. Robustness to Small Size Datasets**

This paper presents a method for estimating brain age using two small datasets, 289 and 580 samples. The architecture is designed to handle limited datasets, making it effective in 3D medical images like MRI or CT. The design considers the three-plane nature of these images, enhancing its applicability and performance. The method also proposes a slice-based 2D CNN divide and conquer method to address data leakage.

**B. Performance on Extended Datasets**

The proposed method, designed for small datasets, shows increased efficacy as the dataset size increases. Experiments show the model's performance remains robust and significantly improves as the dataset grows, highlighting its scalability and versatility. This makes it a flexible solution capable of delivering exceptional performance across diverse dataset sizes.

### C. Scalability of the Model

Our proposed approach incorporates custom backbone structures that are designed to fit the unique characteristics and scale of our dataset. The model's inherent adaptability allows for easy adjustments to accommodate variations in dataset size, thanks to its componentized nature, which allows for seamless transitions between smaller and larger backbones.

### D. Parallelization of the Model

The proposed network introduces a parallelized architecture, deviating from the traditional hierarchical approach, to enhance efficiency and performance by enabling parallel processing of different components. This approach, due to effective componentization, makes parallelization a feasible strategy for concurrent computations, potentially improving the model's overall performance.

### E. Employment of Several 2D CNNs for Age Prediction

The brain structure can be modeled using voxel-based or slice-based approaches. 3D CNNs are commonly used for modeling the entire brain voxel, but this method has challenges due to extensive parameters and the lack of suitable pre-trained deep 3D CNN models for smaller or medium-sized datasets. To address these issues, a different strategy is proposed: using 2D CNNs for slice processing instead of processing the entire voxel simultaneously. This approach incorporates information from all relevant slices of the entire brain, including the axial, coronal, and sagittal planes, by deciding for each part of the brain simultaneously and integrating predictions.

In summary, the key contributions of this paper can be outlined as follows:

- A novel method for brain age estimation based on a dual-stream greedy method is proposed, which simultaneously harnesses the power of state-of-the-art models and can deal with small datasets.

- It is demonstrated that high performance and efficiency can be achieved by our method even with very small datasets, while also being comparable with other state-of-the-art models and extendable for large-scale datasets.

- A comprehensive evaluation and analysis of the method on the IBID dataset and a publicly available dataset called IXI is provided.

- A thorough analysis of each slice and specified brain part was conducted in this study. Their individual contributions to the model's performance were examined, and their unique roles and interdependencies were understood.

The subsequent sections of this paper are structured in the following manner: Section II reviews the related work on brain age estimation and deep learning methods. Section III provides a detailed description of the proposed method called GDSM. Section IV presents the experimental setup, results, and analysis. Section V discusses the limitations and future work of our method. Section VI concludes the paper.

## 2 Related Works

The approaches in brain age estimation can be categorized into three distinct groups called voxel-based and slice-based [6] and region-wise based approaches. An overview of a way of categorization of these approaches is shown Figure 1. The following subsections are about these three types of approaches and are designed concerning this figure. Figure 2 depicts that the number of research papers has evolved over the years and covered by our study, revealing distinct trends in the contributions of 3D, 2D, and 1D perspectives. The line chart reveals temporal patterns in scholarly output and research trajectories in three domains, providing a focused analysis of specific trends within the studied subset.

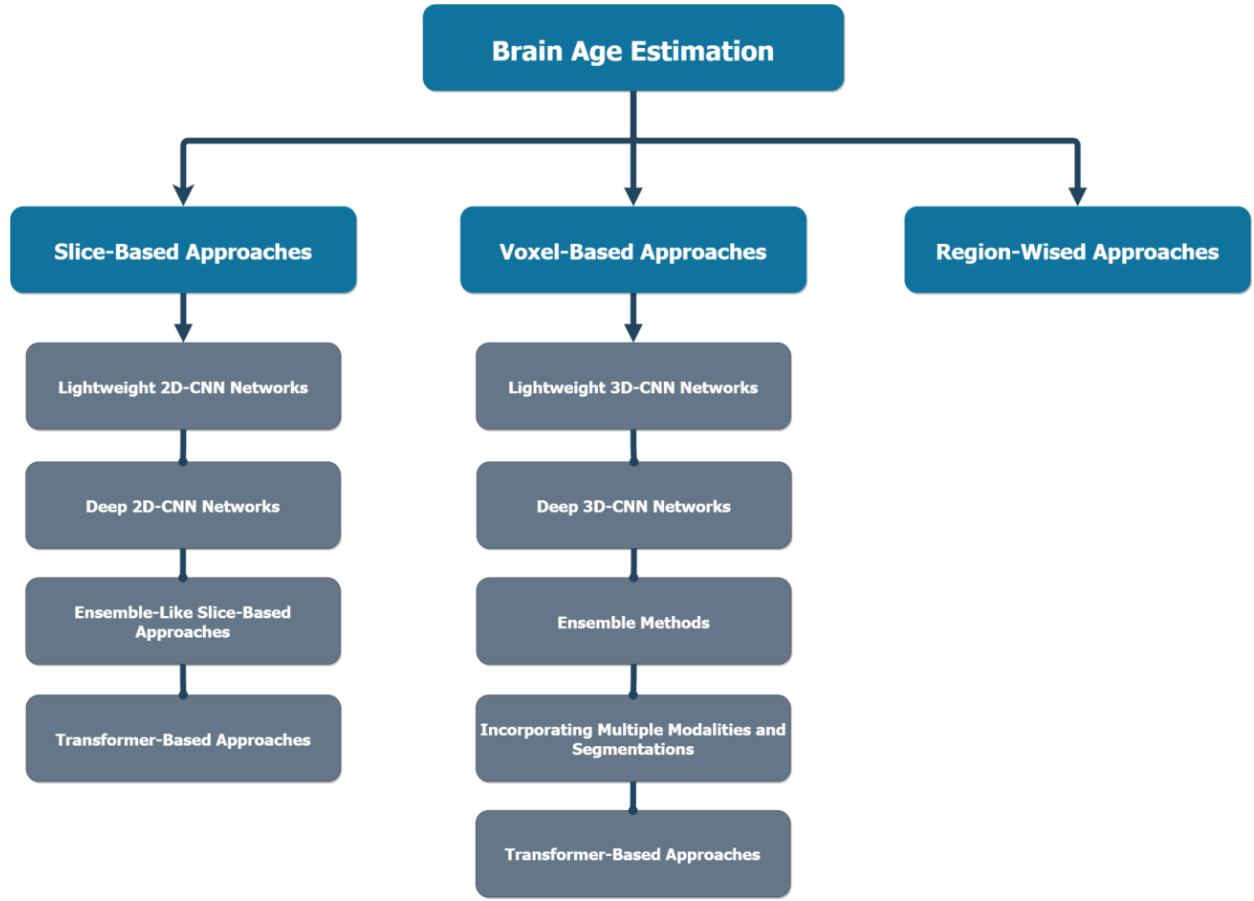

Figure 1- An overview of methodologies for Brain Age Estimation: A comparison between slice-based, voxel-based, and region-wised approaches.

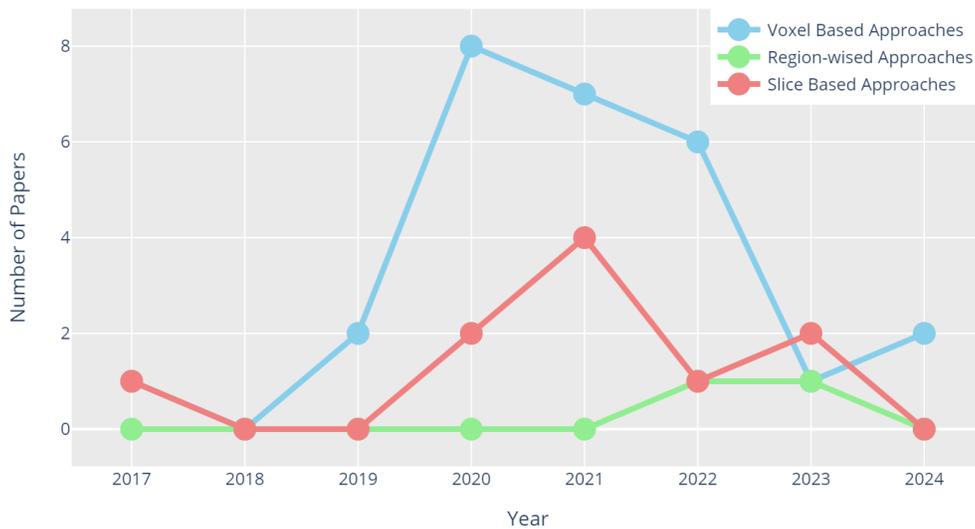

Figure 2- Temporal Distribution of Analyzed Research Papers: Different Perspectives

## 2.1 Voxel-Based Approaches

The first group of the approaches is related to the voxel-based method which uses whole 3D MRI volume for modeling. In most of the proposed approaches in this field, the 3D learnable models were used for backbones, and there were usually multi-site large-scale datasets because 3D models have a massive parameter size and need large-scale datasets. In the remaining part of this section some different methods in this area will be discussed.

### 2.1.1 Utilizing Lightweight Networks

The earliest tries in this area used a lightweight network with a small number of layers. As an earliest examples, Cole et al. [24] proposed a method which firstly gathered White Matter (WM) and Gray Matter (GM) of the brain using SPM12 toolbox and then trained a 3D CNN and demonstrated that GM is better than WM in the brain age estimation. As a drawback, the model is complex and training of it is time-consuming even with a small dataset. In a similar work, Wang et al. [25] segmented WM, GM and Cerebrospinal Fluid (CSF) before architecting model. They used GM for the training of proposed 3D CNN model. In another work, Pardakhti et al. [26] used a 3D CNN followed by a support vector regression (SVR) model and Gaussian process regression (GPR). This study is one of the studies which used only the healthy subjects from IXI dataset. In addition, Hong et al. [27] proposed a 3D CNN network for the brain age estimation in children. They also tested 2D CNNs and demonstrated 3D CNNs' performance is better. In all of the presented works, although the authors tried to propose a neural network which can model brain aging, they posed to problems: (1) The trainable parameters of the models are too much and the computational complexity is too high. (2) They could not use a deep 3D CNN network because of the problem of overfitting. So, they could not extract high-level features from the brain voxels.

### 2.1.2 Utilizing Deeper 3D CNN Networks

Besides simple 3D CNN architecture, 3D versions of other CNN networks such as VGG, Resnet, etc. widely used for brain age estimation. These networks are often expected to perform better than previous networks due to the optimized design. For example, Feng et al. [9], used 5 layers of 3D VGGNet model. Some other works used an ResNet as their model to address age estimation problem. In another work, Jiang et al. [28] segmented MRI images with a reference map into 7 regions. Each segmented areas fed to a 3D VGGNet for the purpose of age estimation. In another endeavor, Dinsdale et al. [29] used 3D VGGNet for brain age estimation on a dataset containing 12,802 structural MRI images from the UK Biobank. The computational complexity of this work is significant and needs much time and resources. Similarly, Peng et al. [30] proposed a simple fully-connected convolutional network similar to VGGNet model. Some details such as dropout, voxel shifting and mirroring. There are some other similar endeavors. For example, Wood et al. [31] and Lee et al. [32] proposed two 3D DenseNet-based models for the brain age prediction. Besides, Fisch et al. [33] presented a 3D CNN network with two ResNet-based layers and a basic CNN block and only T1-weighted pictures were employed.

### 2.1.3 Ensemble Methods

The aforementioned methods employed a singular estimator for each individual subject. To enhance method reliability, some approaches opted for a multi-estimator setup for a single voxel. Although this method led to enhanced outcomes, it also resulted in a rise in computational complexity. The primary challenges in these approaches revolve around model selection and the integration of model results to accurately predict the final age. There are several studies which used ensemble voxel-based techniques. For instance, Kuo et al. [34] presented a sophisticated ensemble approach comprising a model composed of 26 layers of 3D ResNet. Their ensemble strategy encompassed five distinct methods, ending with the use of median aggregation in the final step. In another investigation, Couvy-Duchesne et al. [35] employed an ensemble methodology featuring seven unique estimators. The integration of ensemble model results involved both linear fusion and simple median techniques. Notably, the study revealed superior performance with linear fusion compared to median aggregation. On a different note, Levakov et al. [36] introduced a 3D CNN ensemble method with ten estimators, leveraging linear regression for result integration. Additionally, they implemented data augmentation to augment the dataset's size.

### 2.1.4 Incorporating Multiple Modalities and Segmentation

Despite the various ensemble methods discussed earlier, some approaches differ by incorporating multiple modalities or breaking down one subject's voxel into various segments. While both are sometimes categorized as ensemble methods, a clear distinction exists between them. Ensemble methods typically involve inputting a sample into different models and aggregating their predictions. On the contrary, the following studies entail dividing each sample's image into several distinct parts and employing multiple models for each segment. For instance, in a study by Jonsson et al. [37], more than one modality was utilized. The researchers used WM, GM, Original T1-weighted, and Jacobian map as inputs for multiple 3D ResNet models. In this research, additional information such as scanner details and gender were fed into the last layers of the fully connected neural network (FCNN) to enhance overall performance. In another paper, Hofmann et al. [38] proposed a multi-level ensemble on three different modalities such as T1 weighted, FLAIR, and Susceptibility Weighted Imaging (SWI). The model architecture was relied on 3D CNNs and used linear head model for integrating the results. Similarly in another approach Mouches et al. [39] proposed a two-stream 3D CNN model. In the first stream, they used T1-weighted images and in the second stream, the TOF MRA scans were used as the input. Finally, they applied a linear regression model for the integration of results derived from the two streams.

Some other works used local chunks for brain age estimation. Local chunks for brain age estimation reduce complexity and aid in ROI identification by reducing image dimensions and utilizing aging-influenced parts. For example, Popescu et al. [40] used a 3D U-Net based model which used local parts of brain as input and Bintsi et al. [17] used 3D patches of the brain and 3D ResNet networks for model brain age estimation. This work used Linear Regression and bias correction for the last stage of processing the predicted ages. Using bias correction was a good idea to correct the first predicted ages. Similarly, Kolbeinsson et al. [41] used a 3D ResNet model which input six local parts of the brain including left amygdala, left cerebellar, left insular cortex and the left crus, right hippocampus, and vermis. The other factor which affected by brain aging also assessed. As one of the most recent works, Poloni et al. [42] introduced a novel brain age estimation model comprising two 3D CNN models based on EfficientNet. The input images were 3D voxels of the brain's right and left hippocampus parts. They averaged the results of two networks to calculate the final predicted age.

Beyond the previous research, Numerous studies have focused on the influence of aging on different parts of the brain. For instance, several studies used the right and left hippocampus parts of the brain for brain age estimation [21, 41-45]. These papers analyzed different parts of the brain and demonstrated that the hippocampal region is affected by aging more than others or used the hippocampus region for their downstream classification or regression tasks. The papers under consideration highlight additional brain regions, including the Frontal and Prefrontal Cortex [21, 45], Frontal Operculum [21], Parietal Lateral [21, 45], GM [21, 45], WM [21, 45], CSF [21], and various other regions. These investigations establish correlations between age progression and volumetric alterations in specific brain regions.

### 2.1.5 Transformer based approaches

In the recent years, most of the works used a complex model for the brain age estimation such as vision transformers and attention mechanisms. Transformers were first used for Natural Language Processing

(NLP) tasks [20]. Still, in recent years with the introduction of Vision Transformers (ViT) [46] and its variations, the application of these architectures has been widely extended in image processing tasks including medical imaging. Like most papers in the voxel-based category, these works generally used multi-site datasets to address the problem of overfitting. For an instance, Sheng He et al. [22] proposed a deep relation learning model that took two subjects and calculated 4 relations. 3D CNNs were used as the backbone for extracting feature maps and a transformer was applied to the extracted features for the calculation of four relations comprising arithmetic divide, multiply, maximum, and minimum between two subjects.

Other works used transformers as a part of their model. For example, Zhang et al. [47] proposed Triamese-ViT, which is built with three ViTs, each of which examines 3D images from various angles. A Triamese MLP was also employed in the prediction. In another endeavor, Cai et al. [48] proposed a geometric learning framework for graph transformers to estimate brain age using sMRI and DTI data, using a multi-level graph network and stream convolutional autoencoders.

Some other works, do not use transformers explicitly but use attention mechanisms to enhance the performance or wisely fusion of extracted features in the middle stages of prediction. As an example of these studies, He et al. [49] developed an algorithm that partitions 3D MRI volumes into contrast and morphometry data. They also designed a deep attention-based model with three branches to merge these channels and optimally predict age. In another example, Zhang et al. [50] proposed a model that enhances brain age prediction by integrating anatomical and deep features using an attention-enhanced 3D-CNN. The model used MRI voxels as the input of a VGG-Net-inspired 3D-CNN and an attention network for integrating deep and anatomical features.

These works show their great potential power to model brain age but they suffer from some critical problems. Firstly, they used a multi-site dataset which may pose some problems. Besides, the complexity of their model is too high and they cannot perform well when the training set is a small one. training these models is time-consuming despite the availability of strong resources.

In conclusion for the voxel-based part, due to the nature of brain MRI using whole 3D volumes of the brain for modeling is a good choice as the models capture all spatial and between-slices information but there are some limitations such as lack of data availability and large size of learnable parameters [51] and some authors forced to use multi-site dataset. However, in most of the research, data availability and dataset size are a limitation for most of the researches. Besides, using multi-site datasets may have some problems like scanner effects [52] and social and genetic effects [1] and may not generalized to the subjects from other local geographic areas. For example, working on a dataset comprising Chinese probably cannot work well on a dataset including European subjects.

## 2.2 Slice-Based Approaches

Concerning problems of voxel-based approaches, using slice-based approaches can be useful to address mentioned problems like problems in the complexity of voxel-based approaches [53]. Several works used 2D models to address brain age estimation such as [51, 54-61]. The slice-based approaches can be categorized into several distinct groups in the following parts.

### 2.2.1 Simple and Deep 2D CNN Models

Like the voxel-based section, we start the slice-based model with simple 2D CNN models. As the first example in this area, Huang et al. [54] introduced a 2D CNN network built upon VGGNet. Their approach involved utilizing 15 slices, resulting in initial layers with a channel size of 15. However, a limitation arose as they employed 2D slices from 3D voxels without implementing adequate pre-processing steps, hindering the attainment of optimal results. Additionally, the study lacked the reporting of results from other models.

In another work, Shi et al. [55] proposed a slice-based 2D ResNet that trained on T2-weighted images to estimate the ages of fetal. They also used attention for gathering feature map in the last step of the proposed network. In another vein, Dular et al. [56] used four 2D CNN approach with transfer learning, domain adaptation, and bias correction. In three models they used slices of all 3D voxel for training the models and in one of them they used only 15 axial slices for modeling. While these endeavors aimed to tackle the challenges posed by limited resources and datasets, they fall short in capturing essential between-slices information. This information, distinct from the spatial details captured by 2D CNNs, eludes their grasp, highlighting a gap in their capability to incorporate such crucial data. Some papers proposed a solution for this problem of 2D CNNs. As an example, Lam et al. [51] proposed a model which inputs sequence of slices and encode them using a 2D CNN model. The result feature maps fed to a LSTM model for generating age embedding. They finally applied a linear regression on age embedding to generate final age. Similarly, Lin et al. [57] proposed a 2D AlexNet for the feature extraction of GM part of MRI images and then used principal component analysis (PCA) for feature extraction and finally applied relevance vector machine (RVM) with polynomial kernel as the estimator. In another novel method, Gupta et al. [58] developed a two-component architecture for MRI scans. The first part encodes each 2D slice of an MRI scan, and the second aggregates these encodings. The encoded slices are then processed by an attention-based module and directed to fully connected layers for the final estimation. This architecture is claimed to be more efficient and accurate than conventional 3D-CNN models. In another work, For example, Pilli et al. [61] first segmented voxels into WM, GM and CSF tissues. Then they extracted 2D slices from 3D voxels based on entropy. In the final stage of data preprocessing, they used a Generative Adversarial Network (GAN) as a data augmenter. For the modeling they used a ResNet-50.

### 2.2.2 Ensemble-Like Slice-Based Models

Beside of mentioned approaches, many studies in the slice-based domain, often referred to as ensemble methods by some, may not strictly qualify as ensembles. Instead, they employ various models to process distinct slices, planes, or local patches. For example, Ballester et al. [59] presented a slice-based multi-regressor model incorporating linear regression and CNNs to estimate brain age. The approach involved employing three distinct streams to process age estimation. Each stream was specifically dedicated to a particular plane, and slices from that plane were input into the respective network. Within each stream, the associated slices were fed into the network, and age predictions were made for each slice. The predicted ages from all slices were then integrated using three linear regression models, and the average of the three results was considered as the final predicted age. As another endeavor, Hwang et al. [60] used a 2D CNN for the brain age estimation which inputs multiple slices from MRI voxels. They calculated ages with several slices and then averaged the results. The drawback of this method is that all slices are not influential for the brain age estimation as equal proportion.

### 2.2.3 Transformer based approaches

Like the transformer-based section in voxel-based approaches, there are some papers which used the 2D slices as the input of their model and used transformers in one part of their architecture. As the only example in this part , Sheng He et al. [21] proposed a model that captures both local and global features of the brain using two pathways. The information of local and global parts is integrated using a global-local transformer. Although the computational complexity is reduced compared to similar transformer-based approaches in 3D voxels, there is also a need for high computational resources and big datasets for training these models.

While slice-based approaches may offer advantages over voxel-based methods when faced with data and resource limitations, there are inherent challenges associated with their use. A primary concern with slice-based models is the potential omission of critical information between different slices or planes, a drawback that voxel-based approaches excel in addressing. If the selected slices comprehensively cover aging-related information, a slice-based approach can perform at a comparable level to voxel-based methods. However, assuming the training dataset is sufficiently large, opting for a voxel-based approach may prove superior to a slice-based one. In summary, the problem that specialists should solve is to offer models for different situations especially when the training dataset and computational resources are limited.

## 2.3 Region-wise Approaches

In previous sections, the voxel-based and slice-based approaches were thoroughly analyzed, and some works in each area were reviewed. As mentioned, there is another method of using MRIs for the task of brain age estimation, known as region-wise approaches. In region-wise approaches, before the learning phase, some preprocessing steps are performed to gather the volume of each brain part. The region-wise approach is fundamentally different from the voxel-based and slice-based approaches. Instead of considering the brain as a whole or in slices, this method focuses on specific regions of the brain. These regions are identified based on their anatomical or functional significance in the context of brain aging. The preprocessing steps involve segmenting the brain MRI into different regions using various techniques such as atlas-based or learning-based methods. The volume of each region is then calculated. This data forms the basis for the learning phase. In the learning phase, machine learning algorithms are used to establish a relationship between the volumes of the different brain regions and the age of the individual. The model thus developed can then be used to estimate the brain age of a new individual given their brain MRI. In this study, only a limited number of approaches in this area were analyzed, which did not technically use computer vision techniques. For example, Ahmed et al. [62] first gathered region-wise features and then used some traditional models such as Support Vector Regression (SVR) and Linear Regression for modeling brain age. In another work, Beheshti et al. [63] proposed an algorithm in which voxel-wise and region-wise features were calculated by SPM12 and FreeSurfer toolkits. They finally used the integration of gray matter voxel-wise maps and all region-wise metrics for feeding to an SVR model.

# 3 Material and Methodology
## 3.1 Overview

An overview of the proposed model is depicted in Figure 3. As discussed, the model is justified for small datasets. The model contains three main components. The local stream predicts ages using the local parts of the brain in different planes and slices. There is also a global component that predicts ages using different slices from each of the three planes. These predicted ages from the global and local streams, along with the reference images, are fed into an age correction component. This component takes the predicted ages and tries to enhance the predictions to achieve a better result. This model operates similarly to the global component, with the difference being that it accepts predicted ages as input.

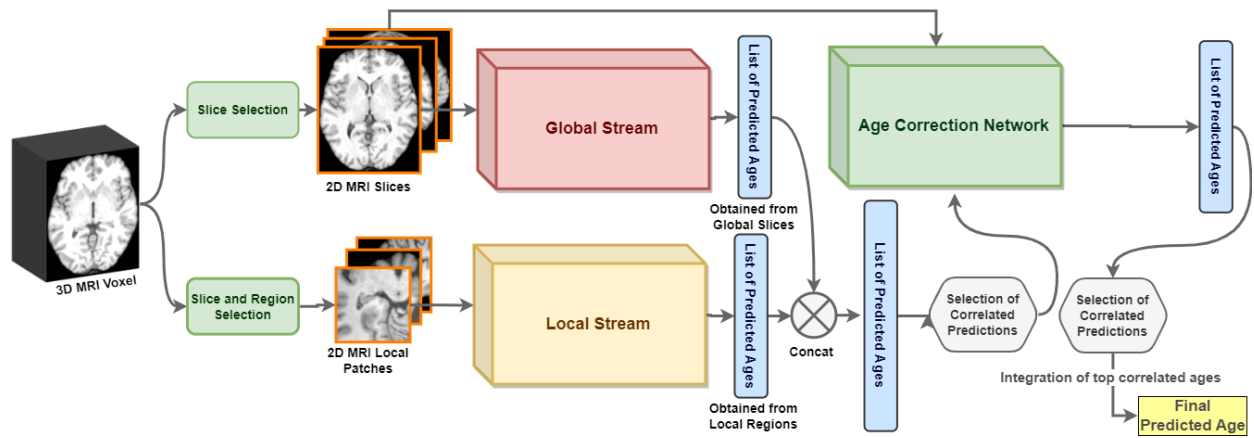

Figure 3 - This is a summary of the architecture used in the study. There are two streams, namely the Local and Global Streams. The predicted ages from these streams are concatenated and fed into the age correlation network.

The second dataset used in this study is the IXI dataset, designed for Information Extraction from Images. It comprises approximately 600 MR images sourced from healthy individuals. The acquisition protocol for each subject's MR image encompasses T1, T2, and PD-weighted images, along with MRA and diffusion-weighted images. This data was gathered from three distinct hospitals in London, utilizing scanners of 1.5T and 3T capacities. The IXI dataset provides demographic information, such as age, gender, and handedness, for each subject. The subjects participating in this dataset are cognitively and mentally healthy and do not have any neurological disorders. Although nearly 600 MR images exist in this dataset, only 565 subjects can be used for the task of brain age estimation because the rest of them do not have any record of their ages.

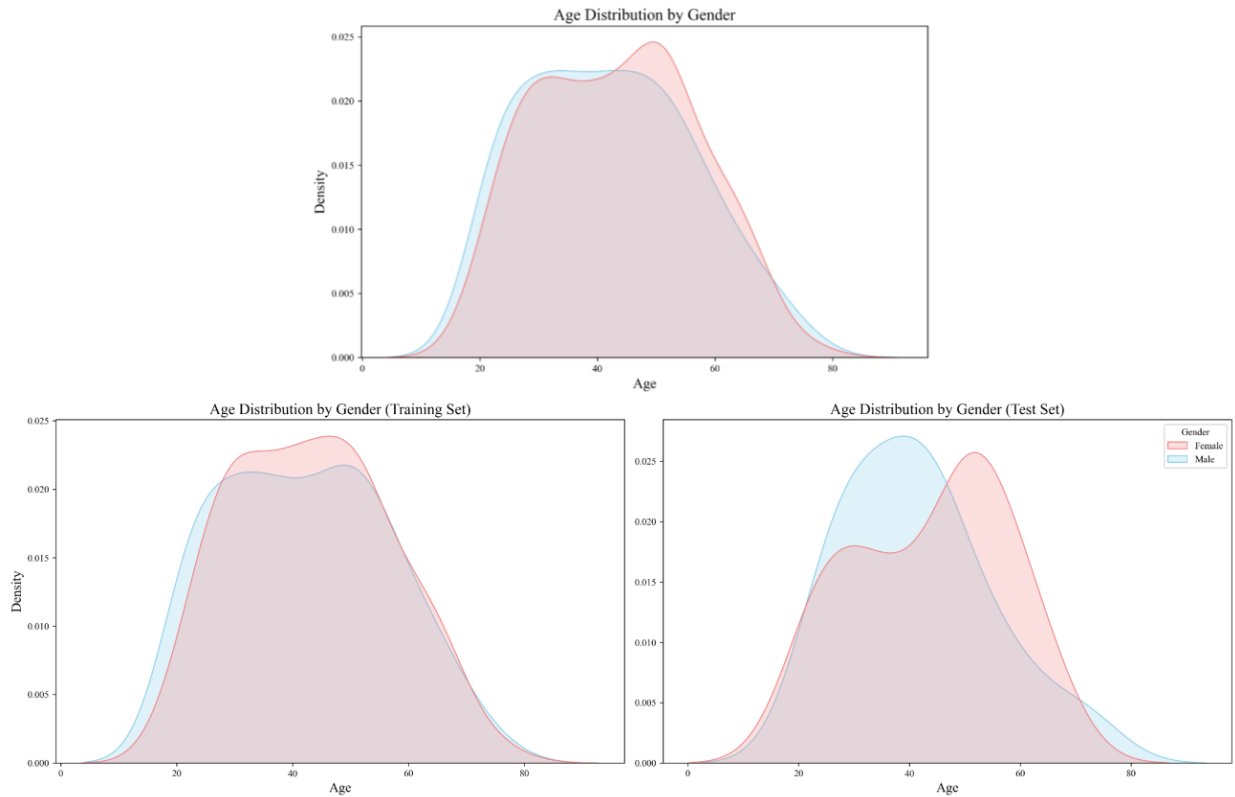

Figure 4 - The Age distribution of IBID dataset by gender before and after splitting.

## 3.2 Data Preprocessing
### 3.2.1 Preprocessing of MRI Voxels

Data preprocessing is an essential phase in the process of estimating brain age using MRI. It involves refining raw MRI data for use in computational age prediction models. This study, like many advanced approaches, uses several preprocessing stages to prepare MRI voxels for model input.

The first step in preprocessing MRI data for brain age estimation is to remove non-relevant elements, such as the skull and surrounding areas, from the original T1-weighted MRI voxels. This is achieved using the SynthStrip [64] model, a U-Net-based deep neural network that was trained on a large dataset to mask the brain tissue. SynthStrip has shown superior performance and better generalization ability across different image types and quality levels compared to other non-learning toolkits.

The second step in preprocessing MRI data is bias correction, which is essential for addressing signal variations and non-uniformities in brain MRI. These irregularities, often due to factors like magnetic field inhomogeneities or radiofrequency coil sensitivity, can distort the data and impact the accuracy of further analyses. Bias correction methods, like the N4 bias correction [65] used in this study, aim to normalize these irregularities to ensure the MRI data accurately reflects the actual anatomical structures.

The T1-weighted images were registered to the MNI152 atlas using the flirt tool in FSL [66], with voxel sizes of 1×1×1mm and 2×2×2mm. This MRI registration process aligns images to a standardized template, reducing spatial variations for consistent comparisons. The final voxel sizes were 182×218×182px for 1mm and 91×109×91px for 2mm.

Patch-scale normalization is used to standardize the intensity values of each 2D training image, scaling them between 0 and 255, and then dividing by 255. This ensures uniformity and addresses intensity inconsistencies. Manual validation is then performed to identify any potential artifacts or abnormalities in the IBID dataset. However, despite identifying some issues, they are not removed due to the small size of the dataset. Some of these problems can be shown in Figure 11.

### 3.2.2 Patch Selection

Patch selection in MRI images is crucial for estimating brain age, as it captures the structural transformations associated with aging. MRI data provides detailed representations of the brain's intricacies, making it a valuable resource for age estimation. The process involves extracting specific regions or sub-regions from scans, focusing on areas susceptible to aging-related changes. This refines the model, simplifies it, and enhances its ability to discern age-related patterns, facilitating generalization across diverse datasets.

This study focuses on selecting specific regions of the brain, such as the hippocampus, frontal cortex, frontal operculum, and parietal lateral, for the local stream of a network. Due to limited dataset availability, findings from other investigations exploring the impact of age on various brain regions were utilized in this study. The relevant articles, benefiting from extensive and reliable datasets, yielded more precise conclusions. Therefore, specific points proposed in those studies have been incorporated into this research. Masks were created using MNI and Harvard Oxford atlases, segmenting brain tissue into specific regions. Patches were resized to 80 by 80 for consistency and divided into right and left halves due to their size. The filling approach was bilinear interpolation. Figure 5 shows an overview of patch selection in this study. While the exploration of suitable local patches has been conducted in the context of brain age estimation tasks previously, delving into a novel area necessitates the assistance of interpretable models such as Grad-CAM [67]. These models can aid researchers in identifying regions of interest, providing valuable insights for focused investigation.

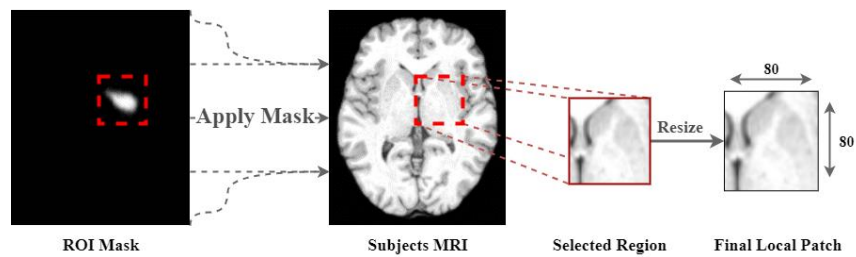

Figure 5 – Selecting approach for the Local ROI with a mask.

### 3.2.3 Slice Selection

The importance of proper slice selection in MRI for accurate age estimation is emphasized. The choice of slices at different angles allows for a comprehensive evaluation of anatomical structures, enhancing the precision of age estimation. Middle slices, which contain a high density of relevant information, are particularly significant [21, 58]. Selecting middle slices can improve efficiency and potentially reduce strain on medical resources. In this work, several slices around the center from each of the three planes - axial, coronal, and sagittal - were selected, while others were omitted. Due to the use of 1mm and 2mm registration of MRI voxels in this study, slice selection varies for the local and global streams. Here is the description for slice selection for local and global stream respectively:

For the local stream, 1mm registration voxels are used due to the smaller sizes of local regions. This choice, driven by the impact of image resolution on model performance, allows for effective analysis of intricate details within these regions, enhancing precision and overall performance. The slice selection for each plane in the local stream is detailed in Table 1.

Table 1- The slice selection for each local parts

| Local Part | Axial Slices Interval | Coronal Slices Interval | Sagittal Slices Interval | Encoded Label |
|---|---|---|---|---|
| Left Hippocampus | [60-79] | [85-120] | [113-128] | 0 |
| Right Hippocampus | [60-79] | [85-120] | [60-72] | 0 |
| Parietal Lateral Left | [66-94] | [80-110] | [100-120] | 1 |
| Parietal Lateral Right | [66-94] | [80-110] | [60-80] | 1 |
| Frontal Opercular Left | [70-84] | [135-140] | [125-140] | 2 |
| Frontal Opercular Right | [70-84] | [135-140] | - | 2 |
| Frontal Lobe Left | [60-94] | - | [100-120] | 3 |
| Frontal Lobe Right | [60-94] | - | [60-80] | 3 |

The global stream is the second stream, where 2D slices are input without patch extraction. To avoid overfitting and reduce the input dimension, a 2mm registration is used for MRI voxels. As mentioned, the dimensions of these registered T1-weighted images are 91×109×91 pixels. The slices around the center, which are more informative and provide clearer details of brain tissue, are chosen. The selected slices are detailed in Table 2.

Table 2- The selected slice numbers for this study are as follows.

| Plane | 1mm MRI slices interval | 2mm MRI slices interval |
|---|---|---|
| Axial | [60-95] | [30-49] |
| Coronal | [80-140] | [40-69] |
| Sagittal | [60-80] ∪ [100-140] | [30-59] |

### 3.3 Data Augmentation

Data augmentation, a technique that applies various transformations to original medical images, plays a pivotal role in enhancing machine-learning models. It artificially enriches the training dataset, thereby improving model performance. However, it requires domain-specific strategies to ensure the clinical relevance of augmented data and to maintain patient data privacy. In this study, we employed a specific augmentation setting for each patch and slice. In the local stream, after the production of each patch in the training dataset preprocessing, augmentation was performed using a random count between 0 and 6. For the global stream, each slice was augmented with a factor, a random number between 0 and 10. Our augmentation setting, inspired by [42], is detailed in Table 3. Additionally, Figure 6 showcases examples of local and global data augmentation. This approach not only enhances the diversity of our dataset but also ensures the integrity of sensitive medical information.

Table 3- Data Augmentation Setting in this study.

| Augmentation Setting | Value |
|---|---|
| Rotation | 20° |
| Width Shift | 0.1 |
| Height Shift | 0.1 |
| Horizontal Flip | True |
| Vertical Flip | True (False for Global pathway) |

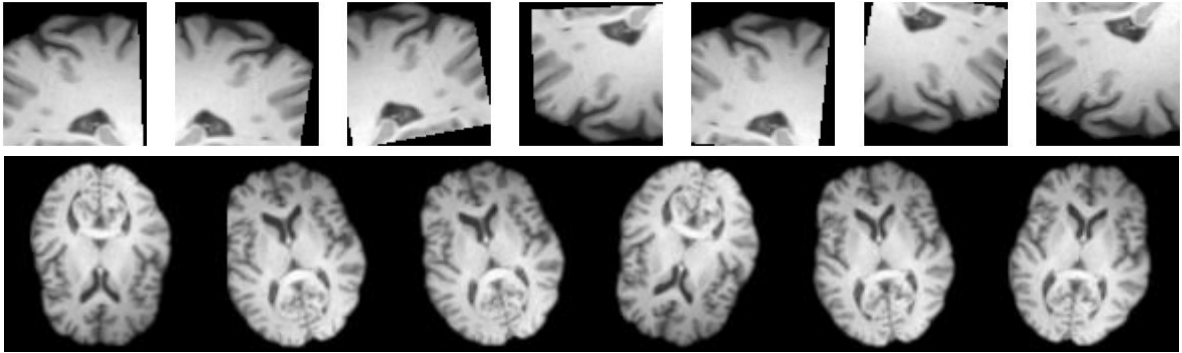

Figure 6- Example of Applying Augmentation on a subject for local and global stream.

## 3.4 Model Architecture
### 3.4.1 Model Overview

As previously described, we propose an architecture that comprises multiple modules and some details. As shown in Figure 3, three main components in this architecture will be explained in detail. These three components are the local stream, the global stream, and the Age Correction component. Initially, an MRI voxel is registered with a 2mm and 1mm registration process. The 1mm registered weighted-T1 images are fed to the local stream, which derives 469 ages per subject. Simultaneously, the global 2mm 2D slices are fed to the global stream to extract the relevant ages. The global stream also outputs 80 predictions for different slices and planes (20 for axial, 30 for coronal, and 30 for sagittal). The culmination of predictions from both streams forms the basis for the final predicted age. Prior to this integration, a distinct component for predicted age correction was introduced. This component utilizes the predicted ages from the local and global streams to estimate new ages relevant to the subjects' brain images. In the following subsections, the independent streams used in the architecture will be described in detail.

### 3.4.2 Local Stream

The architecture of the proposed local stream is shown in

Figure 7. As mentioned in the previous section, the local black box inputs local patches of the brain, which are extracted from MRI voxels by the MNI and Harvard Oxford atlases. The output of this black box is a vector of predicted ages with a size of 469×1. In the black box of the local stream, there are three separate pathways for each plane. The local patches for each plane are fed to the corresponding pathway. For example, the left hippocampus of a slice in the axial plane is given to the local pathway for the axial plane. The reason for using three pathways in the local stream is the high number of variations that the local parts have in each plane. According to Table 1, eight regions have been chosen for local processing. These eight regions for each plane can create a diversity that is hard for a 2D CNN model to handle and can affect performance. Because of this, we decided to split these three pathways to help the models generalize and find a template for aging more easily. Moreover, the number of local training images in all planes was sufficient for training the models. In each pathway, there is a 2D CNN backbone which takes a local patch with a spatial resolution of 80 by 80 pixels. We used a pre-trained Xception [68] model on the ImageNet dataset [69] as the 2D CNN backbone to generate a feature map for each local patch. Xception is a deep CNN architecture that involves depthwise separable convolutions. The core concept of Xception involves substituting the conventional convolution operation with a depthwise separable convolution, which comprises a sequence of a depthwise convolution and a pointwise convolution.

Although the Xception model inputs 2D slices of the brain, the corresponding gender, slice number, and patch type, which come from the encoded label in Table 1, are fed to a fully connected neural network alongside the extracted feature map to predict the ages for each local path. In fact, the estimator uses information on local patches as well as some extra information about slice number, patch type, and gender of a particular subject. For the axial, coronal, and sagittal pathways of the local stream, the output vectors are respectively of size 198×1, 142×1, and 129×1. Finally, these results are concatenated as the output of the local stream black box.

### 3.4.3 Global Stream

The architecture of the proposed global stream is shown in

Figure 7. Unlike the local stream, this stream has only one backbone. This backbone processes 2D slices of the whole brain from each of the three planes: axial, coronal, and sagittal. As shown in Table 2, there are 80 2D images for each subject in total. The backbone processes all the 2D slices and extracts the feature map for each of them. Finally, an FCNN is utilized to estimate the ages using these feature maps. In addition to the feature maps, the corresponding encoded labels for gender, slice number, and plane type, which indicate to which plane the slice belongs, are fed to the FCNN. Clearly, for each subject, the global stream outputs 80 results. Notably, the reason for considering one backbone for the global stream was the limitation in the dataset. In a small dataset, there are not enough independent slices and the slice numbers are limited. Therefore, having three independent backbones for each plane is not logical. However, in the case of having a larger dataset, having three separate pathways is a good choice. The 2D CNN backbone used, which was also pre-trained on the ImageNet dataset, was Xception. This can be changed in different situations depending on the dataset size.

### 3.4.4 Age Correction Component

Enhancing the integration of predicted ages from local and global backbones is a crucial task that offers significant potential for performance enhancement. In this study, our approach involves the use of an age correction component, illustrated in

Figure 7, prior to the integration of ages. This component takes the age results derived from local and global streams as input, forming a vector of size 549, and generates new predicted ages based on 80 global slices. The aim is to produce estimations that surpass the accuracy of previous predictions.

The initial weight of the backbone in this component is set as the learned weights of the global stream's backbone. However, during the training process of the age correction module, several of the last layers of this backbone undergo fine-tuning.

Before the age correction step, we calculate Pearson's correlation to identify which predictions in the set of 549 are more closely related to the actual ages within the training set. This correlation value is computed for each prediction. In the final step, we select the top $C_1$ predictions with the highest correlation coefficients, representing the most correlated predicted ages. These selected predictions are deemed more relevant to the actual age compared to others. Consequently, we construct a predicted age vector for each sample in both the training and test sets, each containing $C_1$ elements.

After calculating all these vectors, the trained global stream's backbone is copied and employed for the age correction task. Fine-tuning is applied to several of the last layers of this backbone, and a new head is created to accept the age vectors instead of the previously discussed information.

The significance of this age correction component is underscored by the imperative that the performance of both global and local streams on the training set closely mirrors the performance on the test set. It is crucial that the predicted ages for the training set serve as a reliable representation of the test set. Even a slight degree of overfitting can have adverse consequences for our model. If the models predict ages well on the training set and do not have any bias to correct, our age correction module does not have anything new to learn. Therefore, if there is any type of bias in the test set, it should also exist in the training set.

### 3.4.5 Final Age aggregation

The final stage of our proposed method involves the aggregation of results. As previously discussed, the age correction component outputs a vector of ages containing 80 elements, similar to the global stream component. In ensemble learning, as seen in previous research, there are various methods to calculate the final age when multiple predictions are available for an individual sample. The simplest method of aggregation is averaging. Other research has used simple regression on the results of the ensemble model. The attention module can also be used for this purpose. In this work, we first calculate the correlation among the 80 predictions and the actual ages. We then select the top $C_2$ correlated estimations. We tested both averaging and regression aggregation approaches for aggregating these $C_2$ estimations. Regression aggregation demonstrated superior performance compared to averaging. However, there were some instances where averaging yielded better results than regression.

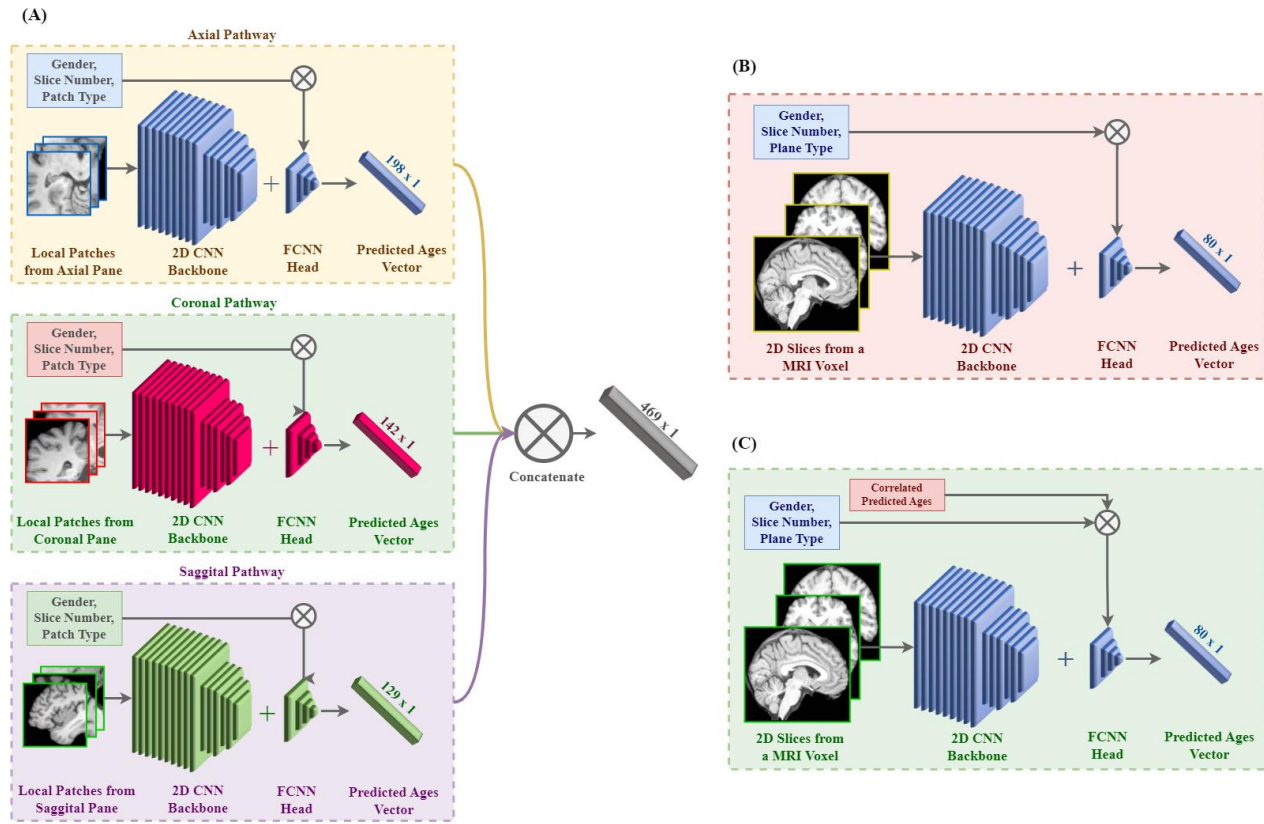

Figure 7- (A) The Local Stream Network, which has three pathways for the axial, coronal, and sagittal planes. (B) The Global Stream Component. (C) The Age Correlation Block.

# 4 Experiments
## 4.1 Evaluation Metrics
### 4.1.1 Mean Absolute Error (MAE)

Mean Absolute Error (MAE) is a crucial tool for assessing the average absolute differences between predicted and actual brain age values, particularly in tasks like brain age estimation. The formula for MAE calculates the mean of the absolute differences for each brain age prediction ($y_i$) and its corresponding actual age ($y_i'$). It is expressed as follows:

$$MAE = \frac{1}{n} \sum_{i=1}^{n} |y_i - y_i'|$$

In this context, $n$ signifies the total count of data points present in the dataset.

### 4.1.2 Mean Square Error (MSE)

Mean Squared Error (MSE) is a regression metric used for brain age estimation. It assesses the average difference between predicted and actual brain age values but is sensitive to outliers.

$$MSE = \frac{1}{n} \sum_{i=1}^{n} (y_i - y_i')^2$$

Compared to MAE, MSE amplifies the impact of larger errors due to the squaring operation. As a result, it may prioritize minimizing extreme errors, potentially at the expense of smaller ones.

### 4.1.3 Root Mean Squared Error (RMSE)

Similar to MSE, Root Mean Squared Error (RMSE) formula is represented as follows:

$$RMSE = \sqrt{\frac{1}{n} \sum_{i=1}^{n} (y_i - y_i')^2}$$

This square root in RMSE transforms the error metric back to the original scale, providing a more interpretable result.

### 4.1.4 Coefficient of Determination (R-squared)

The Coefficient of Determination ($R^2$) is a crucial metric for assessing the effectiveness of a regression model in predicting brain ages. A high $R^2$ value indicates a better fit of the model to the brain age data, explaining 70% of the variability in brain age. However, it should be interpreted alongside other metrics and should take into account limitations such as overfitting and task characteristics. The formula for $R^2$ is as follows:

$$R^2 = \frac{SS_{res}}{SS_{tot}}$$

where $SS_{res}$ is residual sum of squares and defined as follows:

$$SS_{res} = \sum_{i}(y_i - y_i')^2 = \sum_{i} e_i^2$$

And $SS_{tot}$ is total sum of squares (variance of the data) can be defined as follows:

$$SS_{tot} = \sum_{i}(y_i - \bar{y})^2$$

which $\bar{y}$ is the mean of the predicted ages.

### 4.1.5 Pearson Correlation Coefficient

For brain age estimation, the Pearson Correlation Coefficient ($r$) is calculated to assess the linear relationship between the predicted brain ages and the actual brain ages. The formula for $r$ is as follows:

$$r = \frac{\sum_{i=1}^{n}(y_i - \bar{y})(y_i' - \bar{y'})}{\sqrt{\sum_{i=1}^{n}(y_i - \bar{y})^2 \times \sum_{i=1}^{n}(y_i' - \bar{y'})^2}}$$

where $\bar{y'}$ represents the mean of actual ages. The $r$ value, a numerical measure that ranges from -1 to 1, indicates the intensity and direction of a linear correlation between the estimated and true brain ages. A value close to 1 signifies a strong positive correlation, while a value close to -1 implies a strong negative correlation. A value near 0 indicates little to no linear relationship.

## 4.2 Dataset

In this paper, we used two datasets for training and testing. The first dataset, the Iranian Brain Imaging Database (IBID), was conducted in 2017 by Batouli et al. [70]. It comprises 300 neurologically healthy subjects ranging in age from 19 to 77 years. However, eight of these subjects did not have age data, which is crucial for our task. Additionally, three subjects had artifacts in their data, which could bias the training of models. Therefore, we removed these subjects, resulting in a dataset of 289 subjects. To our knowledge, no other papers have worked with this dataset. The main challenge of this dataset is its small size. Figure 4 shows some details of the IBID dataset.

We shuffled this dataset and split it into two groups for training and validation, with a ratio of 80:20. To ensure a fair distribution of samples, we first grouped the samples by age into 20 groups and then used stratified splitting. This approach ensures that our model's performance can generalize across the entire study population. The final sample count was 231 in the training set and 58 in the validation set.

## 4.3 Experimental Settings

We first trained the local component using different local patches for the training step. Several pre-trained deep 2D CNN models were tested to check their performance. In the final stage of backbone testing, we found that the Xception model had the best performance. In addition to testing different backbones, we also tested different hyperparameters. The best hyperparameters are shown in Table 4. These 2D CNN models were tested on the IXI dataset. Although the Xception models were pre-trained on the ImageNet dataset, due to the small size of our dataset, we used the IXI dataset for fine-tuning that pre-trained model. We used the IXI dataset for pre-training our model, which was trained with ImageNet beforehand. The IXI dataset was split into two groups: training and validation, and the hyperparameters were found by evaluating the validation set of the IXI dataset. After training the local backbone with the training set of the IBID dataset, the global backbone was fine-tuned similarly. Then we used a copy of the trained global component for the construction of predicted age correction. In addition, we set $C_1 = 10$, which means the top 10 predicted ages from local and global streams were fed to the age correction module along with the global images. Then we fine-tuned the last layers of this age correction component. For applying the final age aggregation, we set $C_2 = 3$, which means that the top three predicted ages were selected for the final aggregation, and two methods, including averaging and regression methods, were used for final aggregation. Even though in many cases the performance of using regression for final aggregation was better than averaging, the best results were achieved when we used simple averaging.

Table 4- Hyperparameters of proposed method

| Hyperparameter | Value |
|---|---|
| Learning Rate | Schedule Learning Rate (Starting at 0.001) |
| Optimizer | Adam |
| Early Stopping | Yes, Patience = 3 |
| The head architecture | Fully connected; 64:32:16:1 |
| Flattening | Global Average Pooling |
| Number of epochs | 40 |
| Loss | MAE |
| Batch Size | 32 |

The experiments were conducted using a Tesla T4 GPU with 15GB of GPU RAM. However, the available main RAM was only 13GB, which was limited. As a result, the batch size for model training was set to 32.

## 4.4 Comparison With State-of-the-Art Methods

As discussed before, two major approaches stand out in the field of brain age estimation. The first involves employing a specialized model for analyzing 3D voxel data, offering detailed insights into the brain's spatial intricacies. The second approach addresses challenges posed by limited datasets by employing 2D slices and models tailored for individual slices. We present a detailed table encapsulating the results obtained through each methodology to facilitate a nuanced comparison of their outcomes. Table *5* shows a brief performance comparison of different proposed approaches in brain age estimation. This table is sorted by the dataset size in descending order. Clearly, the results cannot be compared to each other because the datasets used in the studies are different. Although some reference datasets like OpenBHB [71] were used, they have not been widely used in recent works. Consequently, to have a fair comparison between the proposed methods, besides the final results, some information about the dataset is also important. For example, dataset size and age range can be two factors that affect the final result. If the dataset size is small, naturally the generalization for the testing loss cannot recede from a particular amount. Additionally, in the case of a small age range, the results can be better compared to a larger age range because of less variation in ages. For example, as shown in Table 5, the MAEs for the UK Biobank are smaller compared to the other datasets. Another fact that can be derived from this table is that with a decrease in the size of datasets, the loss of models increases.

Table 5 - Result comparison of the other proposed methods and our method.

| Paper | Dataset Name | Dataset Size [Age Range] | Voxel/Slice based | Results |
|---|---|---|---|---|
| Wood et al. [31] | Multi-site dataset | 23,302 [19-95] | 3D voxel | MAE= 2.97 |
| Dinsdale et al. [29] | UK Biobank | 19,687 [44.6-84.6] | 3D voxel | MAE= 2.86/3.09, MSE= 13.12/15.13, r= 0.87/0.86 |
| Jonsson et al. [37] | Multi-site dataset | 16,848 [18-80] | 3D voxel | MAE= 3.39 |
| He et al. [49] | Multi-site dataset | 16,705 [0-97] | 3D voxel | MAE= 3.00, r= 0.9840 |
| Cai et al. [48] | UK Biobank, ADNI | 16,458 [46-81] | 3D voxel | MAE= 2.71, RMSE= 3.68 |
| Peng et al. [30] | UK Biobank | 14,503 [44-80] | 3D voxel | MAE= 2.14 |
| Fisch et al. [33] | German National Cohort | 12,864 [20-72] | 3D voxel | MAE= 2.84 ± 0.5 |
| Bintsi et al. [17] | UK Biobank | 13,750 [44-73] | Ensemble 3D patches | MAE = 1.96, r= 0.87 |
| Lam et al. [51] | UK Biobank | 10,446 [45-81] | 2D Slices | MAE= 2.86, RMSE= 3.61 |
| Gupta et al. [58] | UK Biobank | 10,446 [45-81] | Ensemble 2D Slices | MAE= 2.823 |
| Levakov et al. [36] | Multi-site dataset | 10,176 [4-94] | Ensemble 3D voxel | MAE= 3.07, r= 0.98 |
| Feng et al. [9] | Multi-site dataset | 10,158 [18-90] | 3D voxel | MAE= 4.06, r= 0.970 |
| Kolbeinsson et al. [41] | UK Biobank | 9,086 [40–69] | Ensemble 3D patches | MAE = 2.87 |
| He et al. [21] | Multi-site dataset | 8,379 [0-97] | 2D Slices | MAE= 2.70, r= 0.98 |
| He et al. [22] | Multi-site dataset | 6,049 [0-97] | 3D voxel | MAE = 2.38, r= 0.98 |
| Wang et al. [25] | Rotterdam Study | 5,496 [46-96] | 3D voxel | MAE= 4.45±3.59 |
| Ahmed et al. [62] | OpenBHB (Multi-site dataset) | 3965 [6-86] | Region-wise Features | MAE= 3.25, RMSE=4.72, r=0.90 |
| Popescu et al. [40] | Multi-site dataset | 3,463 [18-90] | 3D voxel | MAE= 9.5 |
| Couvy-Duchesne et al. [35] | PAC2019 | 2,640 [17-90] | Ensemble 3D voxel | MAE= 3.33 |
| Ballester et al. [59] | Muti-site PHOTON-AI | 2,639 | Ensemble 2D Slices | MAE= 4.62 |
| Hofmann et al. [38] | LIFE MRI (private) | 2,637 [18-82] | 3D voxel | MAE= 3.37–3.86 |
| Dular et al. [56] | Multi-site dataset | 2,543 [18-96] | 2D Slices | MAE= 3.3, r= 0.91 |
| Zhang et al. [47] | Multi-site dataset | 2,501 [20-94] | 3D voxel | MAE= 2.20, RMSE= 3.26 |
| Kuo et al. [34] | PAC2019 | 2,118 [17-90] | Ensemble 3D voxel | MAE= 3.77, r= 0.90 |
| Mouches et al. [39] | SHIP | 2,074 [21-81] | 3D voxel | MAE = 3.85, r= 0.88 |
| Cole et al. [24] | Brain-Age Normative Control | 2,001 [18-90] | 3D voxel | MAE= 4.16, RMSE= 5.31, r= 0.96 |
| Hwang et al. [60] | Seoul National University Hospital and underwent brain MRI scans | 1,800 [19-89] 560 (IXI as test) [20-86] | Ensemble 2D Slices | MAE = 4.22 (Local)/ 9.96 (IXI) r= 0.862 (local)/ 0.861 (IXI) |
| Poloni et al. [42] | Multi-site dataset | 1,554 [20-85] | Ensemble 3D hippocampus | MAE= 3.64, RMSE= 5.32, r= 0.94 |
| Jiang et al. [28] | Multi-site dataset | 1,454 [18-90] | 3D voxel | MAE= 5.55 |
| Zhang et al. [50] | Multi-site dataset (IXI, ABIDE) | 1,351 [6-90] | 3D voxel | MAE = 3.87, r= 0.93 |
| Huang et al. [54] | Multi-site dataset | 1,099 [20-80] | 2D Slices | MAE= 4.0 |
| Beheshti et al. [63] | Multi-site dataset (IXI, OASIS) | 788 [18-94] | Region-wise Features | MAE= 4.63 |
| Lin et al. [57] | Multi-site dataset | 594 [50-90] | 2D Slices | MAE= 4.51, r= 0.979 |
| Pardakhti et al. [26] | IXI | 562 [20-86] | 3D voxel | MAE= 5.149, RMSE= 13.5 |
| **Proposed Method (GDSM)** | **IBID** | **289 [19-77]** | **2D Slices** | **MAE= 3.25, RMSE= 4.38, r= 0.94** |

## 4.5 Complementary Results

### A. Relation between slice number and model performance

Although we used different slices for estimating brain ages, certain slices of brain aging are significantly impacted by the overall estimation process, shaping a nuanced understanding of the factors influencing brain aging. This highlights the complexity of the prediction process and the need for a nuanced approach that considers the unique contributions of each slice. Understanding the specific characteristics and features of these influential slices is crucial for refined and targeted strategies to enhance the precision and reliability of brain age predictions. Further exploration into the underlying mechanisms governing these influences contributes to the ongoing efforts to unravel the multifaceted nature of the aging brain. Most recent works claimed that the center slices on each plane have the best information about the brain. However, in the task of brain age estimation specifically, it is important to demonstrate which slices are more important for the estimation process. To achieve this goal, we analyzed the performance of our local stream on each plane one by one. Figure 8 shows the impact of each slice from the three planes on the age prediction. Upon observation, it is evident that the middle slices across all three planes are generally more informative. However, an exception is noted in the sagittal plane. Around slice number 40, there is a marked increase in the model's loss. This surge in loss could be attributed to various factors, including potential anatomical variations or anomalies in the brain structure at that specific slice position, or even limitations within the model itself. Upon further investigation, it was determined that this could be a consequence of the nature of the sagittal view, which bisects the brain into left and right halves. Precisely in the few middle slices, the information regarding brain tissue is less abundant compared to other slices in the middle area. Consequently, a temporary increase is observed following a sharp decrease in the loss for the sagittal plane.

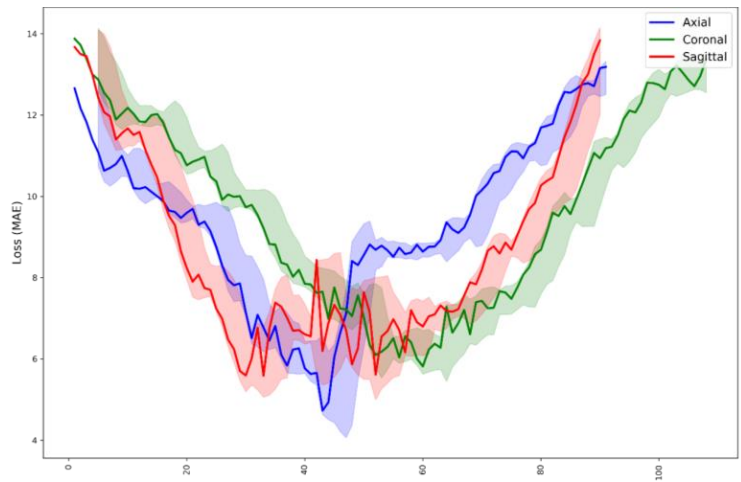

Figure 8- Influence of each slice on the result of the local stream.

### B. Relation between different regions and model performance

Analyzing the relative importance of local brain regions in predicting and influencing outcomes is a crucial aspect of understanding the interplay between neuroscience and machine learning. In recent years, there have been extensive studies into the influence of aging on each part of the brain. Among these parts, as shown in Table 1, we selected the Hippocampus, Parietal Lateral, Frontal Opercular, and Frontal Lobe as the local parts. In this section, we aim to clarify the importance of these local parts and their contribution to the final result. Table 6 shows our model's performance on different regions of the brain. To populate this table, we averaged our model's performance for each brain region on the subjects in the validation set before performing the age correction component. We first chose specific predictions of a region from local stream outputs. Then we selected the top 10 predicted ages from the results and applied an average to these

10 predicted ages. This table highlights the significant importance of each of these brain regions on the final prediction.

Table 6 - Comparison of the impact of each local part of the brain on aging in our study

| Local Part | Validation Loss |
|---|---|
| Hippocampus | 4.9733 |
| Parietal Lateral | 4.8322 |
| Frontal Opercular | 5.3502 |
| Frontal Lobe | 4.9556 |
| All Regions | 4.7242 |

## C. Component Influence Analysis (Ablation Study)

Analyzing the impact of each component on the final architecture is crucial, given the diverse elements utilized in its construction. This analysis is not merely an academic exercise, but a necessary step in understanding the interplay of the components and their collective influence on the overall system. As we have thoroughly explained, there were three main components named Local, Global, and Age correction in our proposed model. Each of these components plays a unique role in the system. The Local component focuses on the local influential parts of the brain on aging, the Global component provides a holistic view, and the Age correction component enhances the quality of prediction with respect to the global view. The strength of our model lies in the synergy of these three modules. As shown in Table 7, the utilization of these three modules together had the best performance among the other combinations. This result underscores the importance of considering all three components in tandem rather than in isolation.

In addition to the analysis of the impact of each module in our proposed approach, the average and regression procedures were taken as the final integration process for predicted ages. Table 7 also shows the importance of using regression integration. To complete Table 7, we used the top 10 correlated predicted ages from the results of each group. Subsequently, simple average integration and various traditional regression models, including linear regression, support vector regression, and random forest regression, were evaluated. Among these, the support vector regressor emerged as the most effective. Although Table 7 shows that the regression integration overshadows simple averaging, the best performance for our model was achieved by using average integration of only the top 3 correlated predicted ages after performing age correction. However, in that case, the performance of averaging and regression did not show a significant difference.

Table 7- Ablation study of our proposed network

| Model | Validation Loss | |
|---|---|---|
| | Average Integration | Regression Integration |
| Local Stream | 4.7219 | 4.7618 |
| Global Stream | 4.3542 | 4.2159 |
| Local + Global | 4.2633 | 3.9142 |
| **Local + Global + Age Correction** | **3.3359** | *3.3079* |

## D. Generalizability of the proposed model on a similar dataset

To address concerns about the generalization of our model beyond the IBID dataset and small-scale scenarios, we conducted additional experiments on the IXI dataset, another dataset characterized by its relatively small size. This deliberate extension of our testing to a different dataset allows us to evaluate the adaptability and performance of our model in a varied context. By including IXI in our analysis, we aim to demonstrate the robustness of our approach and provide insights into the potential applicability of our model across diverse datasets with similar characteristics. The results obtained from the IXI dataset will be discussed in the subsequent sections, shedding light on the generalization capabilities and extending the scope of our proposed methodology. To achieve this goal, we split the IXI dataset into a 75-25 proportion for training and testing, respectively. We employed some pre-trained 2D-CNN network (VGG16) on the ImageNet and fine-tuned it using the IXI training set.

The performance of our model on the IXI dataset is presented in Table 8, alongside the results of other models that exclusively utilized the IXI dataset for evaluation. This comparative analysis provides insights into how our approach performs in the context of similar datasets and allows for a direct comparison with existing models that have been tailored to the characteristics of the IXI dataset.

Table 8- Compare the results of models that used only the IXI dataset for the training process.

| Model | MAE | Other measures |
|---|---|---|
| Keshavarz et al. [72] | - | RMSE= 11.32 |
| Pardakhti et al. [73] | 5.813 | RMSE= 8.641 |
| Pardakhti et al. [26] | 5.149 | RMSE = 13.5 |
| Mishra et al. [74] | 4.61 | RMSE = 6.5, r= 0.91 |
| **GDSM (Ours)** | **4.1805** | **RMSE = 5.53, r = 0.94** |

# 5 Discussion

In this work, we proposed a model for the task of brain age estimation, particularly when the dataset size is limited. A slice-based approach was proposed, which, instead of modeling the whole voxel at once, outputs an age for each local and global part of the brain voxel. Although the majority of slice-based approaches miss the information between the slices, our approach used all the informative slices separately and then integrated the results. Given that our model incorporated a global-local architecture, significant focus was directed towards the local regions of the brain associated with aging. This architecture achieved spectacular performance on a small Iranian brain dataset. It achieved a MAE of 3.25 years, an RMSE of 4.38 years, and an r coefficient of 0.94. The correlation plot of our proposed model on the validation set of the IBID dataset is shown in Figure 9.

To demonstrate the biases that exist in predictions for the validation set, we analyzed two subjects in Figure 10. Although the chronological ages of the subjects were similar, we can visibly discern that the volume of specific parts of their brains differed. This fact illustrates that the concept of 'healthiness' is not binary but relative, encompassing a variety of subjects within a specific age group [29]. For instance, a person deemed healthy in one age group could also be considered healthy in another age group, whether older or younger. Our approach uses multiple references for the final prediction, reducing the probability of errors compared to using just one estimator or a single region. Furthermore, due to multiple estimations for each subject, we can estimate a possible age range for each subject, within which a person with that brain could be interpreted as healthy.

In addition to these challenges, there were other problems in the dataset that could lead to model bias. For example, as shown in Figure 11, some subjects could visually be interpreted as 'unhealthy' due to significant differences compared to their age group. The image shows that the brain of a person aged 58 is much more decomposed from aging compared to a 'healthy' person from the same or even older age groups. The presence of these types of problems in our models' training set can cause significant issues due to the limited size of the dataset. In small datasets, the existence of each artifact can cause a significant bias and hinder the generalization of our model. Therefore, it is crucial for our model, and other models designed for small datasets, to thoroughly check all training data and ensure data validation.

In another experiment, to demonstrate the robust performance across all small datasets, this approach was tested on the IXI dataset. It achieved state-of-the-art performance with an MAE of 4.1805, an RMSE of 5.53, and an r coefficient of 0.94. These findings show competitiveness with results reported in the literature for previous methods, as shown in Table *5*, even when only slices were used instead of the whole voxel. For larger datasets, such as multisite datasets, this approach can perform well with minor changes in the architecture or data preprocessing.

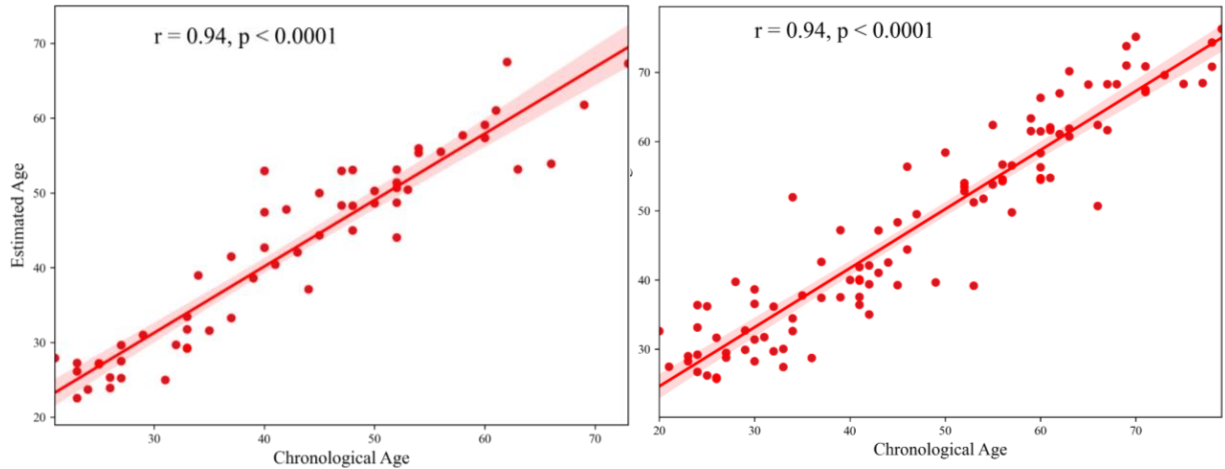

Figure 9- Correlation plot of our proposed model on the validation data of IBID (left) and IXI (right).

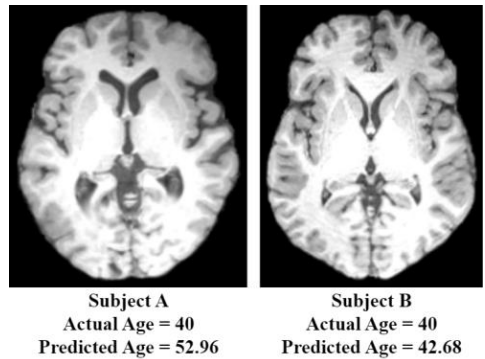

Figure 10- A comparison of the MR images of two subjects in the validation set, both with a chronological age of 40.

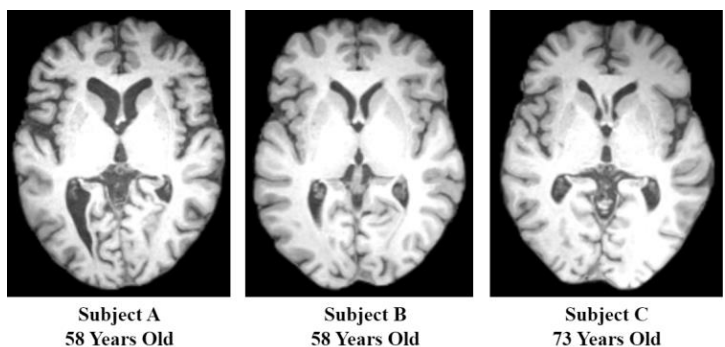

Figure 11- An example of existing problems in the database. Both Subject A and Subject B are 58 years old, yet there is a substantial discrepancy in their respective volumes. It appears that Subject A's MRI contains some artifacts. Upon comparing subjects A and C, it becomes evident that the rate of deterioration in Subject A significantly surpasses that of Subject C with a chronological age of 73.

# 6 Conclusion

The existing body of literature has introduced several deep-learning models capable of precisely estimating the age of the brain in individuals without neurological disorders. Large datasets were required to train the massive models used in the majority of these approaches because they had a large number of parameters to train. Despite attempts to aggregate smaller publicly available datasets to create a substantial dataset, these models struggle to perform optimally with limited data. Our study was motivated by the necessity to tackle scenarios where a considerable dataset for training is unavailable. In the context of studies focusing on specific groups or regions especially, the issue of dataset leakage becomes apparent. Our research endeavors to mitigate these concerns by addressing the limitations associated with dataset size, ensuring that our models maintain robust performance even when dataset leakage is a potential risk. This methodology strengthens the validity of our results and highlights the ethical issues that are important when conducting research with certain populations or geographical settings. In this study, we proposed a slice-based deep local-global network to tackle this limitation. The final predicted age is derived from some sub-predictions from local and global streams. We finally used a bias correction for the predicted ages and because of that, we reached an extraordinary MAE and RMSE for small datasets and highly competitive results for larger datasets.

As with any novel approach, there are opportunities for further enhancement and exploration. Future research endeavors could delve into optimizing the slice and patch selection procedure by introducing a new automated approach based on attention-based selection approaches. Besides, this work can be extended to different small datasets and tested on a multi-site dataset to show its performance on large datasets. The versatility of this architecture makes it applicable not only to neurology tasks but also to other medical images, showcasing its potential for a range of tasks. Moreover, developing a system for finding invalid data and outliers for the task of brain age estimation is a crucial task.

# 7 Data availability

The IXI dataset is publicly available at https://brain-development.org/ixi-dataset/ and the IBID dataset can be obtained from the corresponding author, Amirhossein Batouli (batouli@sina.tums.ac.ir).

# 8 Acknowledgments

This work did not receive financial support from any funding agency.

# 9 CRediT authorship contribution statement

**Iman Kianian:** Conceptualization, Methodology, Software, Writing original draft.

**Hedieh Sajedi:** Supervision, Validation, Review & editing

**Conflict of Interest**

The authors declare no commercial interests with the content of this paper.